\documentclass{article}
\usepackage{cite}
\usepackage{spconf,amsmath,amssymb,amsfonts}
\usepackage{algorithmic}
\usepackage{graphicx}
\usepackage{psfrag}
\usepackage{textcomp}
\usepackage{xcolor}
\usepackage{url}
\usepackage[absolute]{textpos}
\newcommand{\newBDnameLong}{Akima Bj{\o}ntegaard Delta (ABD)}
\newcommand{\newBDnameShort}{ABD}

\newcommand{\copyrightstatement}{
    \begin{textblock}{15}(0.5,0.3)    % tweak here: {box width}(leftposition, rightposition)
         \noindent
         \centering
         \textblockcolour{white}
         \footnotesize
         \copyright 2022 IEEE. Personal use of this material is permitted. Permission from IEEE must be obtained for all other uses, in any current or future media, including reprinting/republishing this material for advertising or promotional purposes, creating new collective works, for resale or redistribution to servers or lists, or reuse of any copyrighted component of this work in other works
    \end{textblock}
}

\def\BibTeX{{\rm B\kern-.05em{\sc i\kern-.025em b}\kern-.08em
    T\kern-.1667em\lower.7ex\hbox{E}\kern-.125emX}}
\begin{document}

\copyrightstatement

\title{BEYOND BJ{\O}NTEGAARD: \\ LIMITS OF VIDEO COMPRESSION PERFORMANCE COMPARISONS }

\name{Christian Herglotz, Matthias Kr\"anzler, Ruben Mons, Andr\'e Kaup}
\address{{Multimedia Communications and Signal Processing} \\
{Friedrich-Alexander University Erlangen-N\"urnberg (FAU)}\\
Erlangen, Germany \\
\{christian.herglotz,matthias.kraenzler, ruben.mons,  andre.kaup\}@fau.de}

\maketitle

%\IEEEpubid{\makebox[\columnwidth]{978-1-7281-9320-5/20/\$31.00 \copyright 2020 IEEE \hfill} \hspace{\columnsep}\makebox[\columnwidth]{ }}
%\IEEEpubid{\begin{minipage}[t]{\textwidth}\ \\[10pt] 978-1-7281-9320-5/20/\$31.00~\copyright~2020 IEEE\end{minipage}}

\begin{abstract}
For $20$ years, the gold standard to evaluate the performance of video codecs is to calculate average differences between rate-distortion curves, also called the ``Bj{\o}ntegaard Delta''. With the help of this tool, the compression performance of codecs can be compared. In the past years, we could observe that the calculus was also deployed for other metrics than bitrate and distortion in terms of peak signal-to-noise ratio, for example other quality metrics such as video multi-method assessment fusion or hardware-dependent metrics such as the decoding energy. However, it is unclear whether the Bj{\o}ntegaard Delta is a valid way to evaluate these metrics. To this end, this paper reviews several interpolation methods and evaluates their accuracy using different performance metrics. As a result, we propose to use a novel approach based on Akima interpolation, which returns  the most accurate results for a large variety of performance metrics. The approximation accuracy of this new method is determined to be below a bound of $1.5\%$. %, its results are mathematically not valid. 
\end{abstract}


\section{Introduction}
\label{sec:intro}
In the past twenty years, video communication has become an integral part of the daily lives of people all around the world. Nowadays, more than three quarters of the global Internet traffic is composed of video data \cite{cisco20}, which underlines the importance of this technology. At the same time, the quality of the video data increased significantly due to ultra high definition resolutions, higher frame rates, and dynamic ranges, which leads to even larger bitrates required for transmission. As a consequence, more sophisticated and powerful compression technologies were developed to mitigate the burden on storage and transmission. As such, every couple of years, new codecs such as H.264/AVC, HEVC, and VVC were developed, which all aimed at reducing the bitrate at the same visual quality by a factor of two with respect to its predecessor \cite{Bross21}. 

\begin{figure}[t]
\centering
\psfrag{014}[tc][tc]{ Rate in Mbps}%
\psfrag{015}[b][t]{ PSNR in dB}%
\providecommand\matlabtextB{\color[rgb]{0.150,0.150,0.150}}%
\psfrag{000}[tc][tc]{}%
\psfrag{001}[ct][ct]{\matlabtextB $0$}%
\psfrag{002}[ct][ct]{\matlabtextB $5$}%
\psfrag{003}[ct][ct]{\matlabtextB $10$}%
\psfrag{004}[ct][ct]{\matlabtextB $15$}%
\psfrag{005}[ct][ct]{\matlabtextB $20$}%
\psfrag{006}[rc][rc]{\matlabtextB $33$}%
\psfrag{007}[rc][rc]{\matlabtextB $34$}%
\psfrag{008}[rc][rc]{\matlabtextB $35$}%
\psfrag{009}[rc][rc]{\matlabtextB $36$}%
\psfrag{010}[rc][rc]{\matlabtextB $37$}%
\psfrag{011}[rc][rc]{\matlabtextB $38$}%
\psfrag{012}[rc][rc]{\matlabtextB $39$}%
\psfrag{013}[rc][rc]{\matlabtextB $40$}%
\psfrag{Codec A}[l][l]{Codec A}%
\psfrag{Codec B}[l][l]{Codec B}%
\psfrag{bounds}[l][l]{Bounds}%
\psfrag{Integration Areaaaaaa}[l][l]{Integration area}%
\includegraphics[width=.42\textwidth]{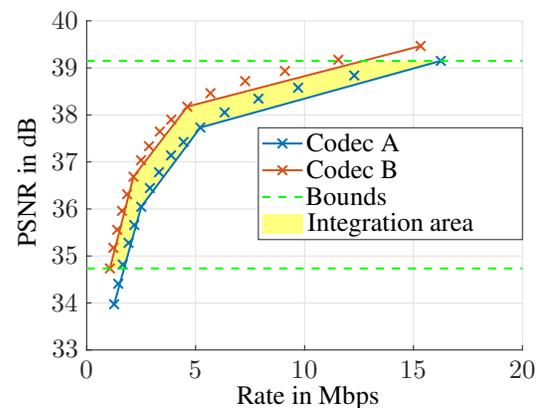} 
\vspace{-.4cm}
\caption{Example for rate-distortion points (x'es) of two different codecs (blue and red), interpolated rate-distortion curves (straight lines), integration bounds for the calculation of average bitrate differences (green dashed lines), and integration area (yellow surface). }
\label{fig:BD_example}
\vspace{-.3cm}
\end{figure}

To this end, a metric was needed that allows to measure the performance of a new compression algorithm. For this purpose, an algorithm known as Bj{\o}ntegaard Delta (BD) was proposed in \cite{Bjonte01}, which calculates the average difference between two curves in a rate-distortion diagram as illustrated in Fig.~\ref{fig:BD_example}. This algorithm can be used in two different ways: Either, the mean vertical difference between the two curves is calculated resulting in an average difference in the quality metric, or the mean horizontal distance is computed, which leads to an average bitrate difference. In the latter case, the relative difference is frequently calculated as a percentage number. The latter approach was the main tool used in standardization \cite{JVET-S2005,Strom21} because the main goal of compression is to reduce bitrates at constant qualities. Hence, we focus our work on this variant. %, which is illustrated in Fig.~\ref{fig:BD_example}. 

In \cite{Bjonte01}, the BD calculus was designed to compare curves representing the relation between the rate in terms of transmission bitrate (e.g., in kilobytes per second) and the distortion in terms of the peak signal-to-noise ratio (PSNR), which was also the main target in standardization \cite{JVET-S2005}. %,JCTVC-J1000}. 
 However, in the last several years, other metrics attracted more attention. A major reason is that the PSNR has a relatively low correlation with subjective quality when evaluating across different codecs, as reported in \cite{Huynh08}. As a consequence, new quality metrics such as the structural similarity index (SSIM) \cite{Wang04} or the video multi-method assessment fusion (VMAF) \cite{Bampis18} were developed, which aim to have a higher correlation with the subjective quality of videos. Consequently, researchers started evaluating the performance of new compression algorithms replacing the classic PSNR in rate-distortion curves with these new quality metrics \cite{Mercat21,Fischer20}. %\cite{Mercat21,Sharabayko16,Herrou16,Fischer20}. 
Furthermore, results from subjective testing were also employed replacing the PSNR with mean-opinion scores (MOS)  \cite{Ohm12,Bross21}. % \cite{Ohm12,Tan16,Bross21}. 

%For object detection applications, another quality metric was developed called mean average precision (mAP) \cite{Cordts16}. This metric was also used extensively to calculate BD-rate differences between video codecs \cite{Fischer20b,Choi18,Fischer20c}, where the quality was expressed in terms of detection accuracy of a subsequent object detection network. 
Furthermore, in contrast to replacing the quality metric, it was also proposed to replace the bitrate by other metrics describing technically limiting factors, for example the decoding energy or the decoding time \cite{Herglotz19,Herglotz20b,Kraenzler20}.

For all these examples, the classic BD calculus was used to calculate the compression performance of video codecs with respect to different quality or technical metrics, which we call performance metrics (PMs) in the following. However, to the best of our knowledge, it was never studied whether the BD calculus is a feasible, correct, and accurate way to compare codecs. To fill this gap, we perform an in-depth analysis of the BD calculus in this paper, with a particular focus on the interpolation algorithms, different performance metrics, and a representative set of sequences. As a result, we propose a novel method to calculate BD values based on Akima interpolation, which we call \newBDnameLong. %We find that for all PMs, Akima interpolation is the most accurate way to interpolate PM points. 

This paper is organized as follows. First, Section~\ref{sec:BD} reviews the BD-rate calculation formula. % and points out possible sources of inaccuracy. 
Afterwards, Section~\ref{sec:acc} presents a method to evaluate the accuracy of the BD calculus. Then, Section~\ref{sec:eval} introduces our evaluation setup and Section~\ref{secsec:evalInterp} provides explicit error values that can be expected when applying the BD calculus on any kind of PM. Finally, Section~\ref{sec:concl} concludes this paper.

%Section~\ref{sec:BD} reviews the BD-rate calculation formula and points out possible sources of inaccuracy. Afterwards, Section~\ref{sec:metrics} provides an overview of the metrics tested in this paper. Then, Section~\ref{sec:eval} introduces our evaluation set and provides explicit error values that can be expected when applying the BD calculus on any kind of metric. Finally, Section~\ref{sec:concl} concludes this paper. 

\section{The Bj{\O}ntegaard Delta-Rate Calculation}
\label{sec:BD}
In \cite{Bjonte01}, the BD-rate is calculated on rate-distortion curves and is obtained by the following procedure. First, four different rate points or target qualities $i\in\{1,2,3,4\}$ are chosen for a certain input sequence $s$. For these four points, bit streams are encoded with two different codecs $k\in\{\mathrm{A},\mathrm{B}\}$. The actual bitrate $R_{k,i,s}$ and the actual distortion $D_{k,i,s}$ in terms of PSNR for the resulting eight bit streams is then used as a basis and exemplarily depicted as `x'-markers in Fig.~\ref{fig:BD_example}. Without loss of generality, we define the PSNR values to be monotonically increasing with $D_{k,1,s} < D_{k,2,s} < D_{k,3,s} < D_{k,4,s}$. As a consequence, the corresponding bitrates $R_{k,i,s}$ are monotonically increasing, too.

Then, to take into account that the bitrate can vary by multiple orders of magnitude, the bitrate $R_{k,i,s}$ is converted to the logarithmic domain with 
\begin{equation}
r_{k,i,s} = \log_{10}(R_{k,i,s}).
\label{eq:logR}
\end{equation}
This step is performed to ensure that mean BD-rate values are not biased towards higher bitrates \cite{Bjonte01}. 

In the next step, using an interpolation function, polynomial curves are fitted for both codecs $k$ which use the marker positions $\{D_{k,i,s},r_{k,i,s}\}$ as supporting points. % resulting in the solid lines in Fig.~\ref{fig:BD_example}. 
As four supporting points are used, the interpolated curve is represented by a third order polynomial of the form 
\begin{equation}
\hat r_{k,s}(D%_{k,s}
) = a_{k,s} + b_{k,s}\cdot D%_{k,s}
 + c_{k,s}\cdot D%_{k,s}
 ^2 + d_{k,s}\cdot D%_{k,s}
 ^3, 
\label{eq:BDpoly}
\end{equation}
where the parameters $a_{k,s}$, $b_{k,s}$, $c_{k,s}$, and $d_{k,s}$ are derived by interpolation for both codecs. Interpolation is performed in such a way that the resulting curves pass through all supporting points $\{D_{k,i,s},r_{k,i,s}\}$.

%TODO: Interpolation methods are ... Either explain here or later. 

Afterwards, the average difference between the two resulting curves is calculated by integration. As upper and lower bound of integration, the overlapping parts of the two curves are selected, which are calculated by 
\begin{align}
D_{s,\mathrm{low}} = & \max\left( D_{\mathrm{A},1,s},D_{\mathrm{B},1,s}\right) \label{eq:lowBound} \\
D_{s,\mathrm{high}} = & \min\left( D_{\mathrm{A},4,s},D_{\mathrm{B},4,s}\right) \label{eq:highBound}. 
\end{align}
In Fig.~\ref{fig:BD_example}, these bounds are indicated by dashed lines. 

Finally, the average relative difference between the two curves, which is the BD-rate $\Delta R_s$, is calculated by the following integration
%\begin{equation}
%\Delta R = \frac{1}{D_\mathrm{high}-D_\mathrm{low}} \int_{D_\mathrm{low}}^{D_\mathrm{high}}
% \frac{\hat r_\mathrm{B}(D_\mathrm{B})-\hat r_\mathrm{A}(D_\mathrm{A})}{\hat r_\mathrm{A}(D_\mathrm{A})}\mathrm{d}D, \label{eq:bdRateInt}
%\end{equation}
%\begin{align}
%\Delta R =&  \frac{1}{D_\mathrm{high}-D_\mathrm{low}} \int_{D_\mathrm{low}}^{D_\mathrm{high}}
% \frac{\hat R_\mathrm{B}(D_\mathrm{B})-\hat R_\mathrm{A}(D_\mathrm{A})}{\hat R_\mathrm{A}(D_\mathrm{A})}\mathrm{d}D \label{eq:bdRateInt}\\
% = & \frac{1}{D_\mathrm{high}-D_\mathrm{low}} \int_{D_\mathrm{low}}^{D_\mathrm{high}}
% \frac{\hat R_\mathrm{B}(D_\mathrm{B})}{\hat R_\mathrm{A}(D_\mathrm{A})}-1\mathrm{d}D \\
% = & \frac{1}{D_\mathrm{high}-D_\mathrm{low}} \int_{D_\mathrm{low}}^{D_\mathrm{high}}
% \frac{10^{\hat r_\mathrm{B}(D_\mathrm{B})}}{10^{\hat r_\mathrm{A}(D_\mathrm{A})}}\mathrm{d}D -1\\
% = & \frac{1}{D_\mathrm{high}-D_\mathrm{low}} \int_{D_\mathrm{low}}^{D_\mathrm{high}}
% 10^{\hat r_\mathrm{B}(D_\mathrm{B})- \hat r_\mathrm{A}(D_\mathrm{A})}\mathrm{d}D -1\\
% = & 10^{\frac{1}{D_\mathrm{high}-D_\mathrm{low}} \int_{D_\mathrm{low}}^{D_\mathrm{high}}
% \hat r_\mathrm{B}(D_\mathrm{B})- \hat r_\mathrm{A}(D_\mathrm{A})\mathrm{d}D} -1\\
%\end{align}
\begin{equation}
\Delta R_s =  10^{\frac{1}{D_{s,\mathrm{high}}-D_{s,\mathrm{low}}} \int_{D_{s,\mathrm{low}}}^{D_{s,\mathrm{high}}}
 \hat r_{\mathrm{B},s}(D%_{\mathrm{B},s}
 )
 - \hat r_{\mathrm{A},s}(D%_{\mathrm{A},s}
 )\mathrm{d}D} -1, \label{eq:bdRateInt}
\end{equation}
which describes the relative bitrate difference in the bounds of \eqref{eq:lowBound} and \eqref{eq:highBound} of sequence $s$ when encoded with  codec $k=\mathrm{B}$ with respect to codec $k=\mathrm{A}$. 

%\section{Evaluation Metrics for Video Compression}
%\label{sec:metrics}
%As mentioned before, the ``traditional'' metrics for the calculation of the BD-rate are the logarithm of the bitrate and the PSNR. However, other metrics

\section{ BD-Rate Assessment}
\label{sec:acc}
The idea of the BD-rate calculus is to allow the accurate calculation of average differences between RD curves, where usually only discrete points are available in the RD space. 
Hence, it was proposed to interpolate the available points with a polynomial \eqref{eq:BDpoly} and use an integration to get the average difference between the continuous curves \cite{Bjonte01}. 

Still, this idea leads to an important question: Are the resulting continuous curves representative for the true RD behavior of the two codecs? The interpolation method only ensures that the four input RD points $\{D_{k,i,s},r_{k,i,s}\}$ are located on the interpolated curve. In contrast, it is unclear, and, to the best of our knowledge, it was never analyzed in detail whether the curves are representative for RD points between the supporting points. 

As a consequence, we propose to evaluate the BD calculus on all available RD points in the common region of interest. For many encoders, these RD points can be generated by encoding at different quantization parameters (QPs) \cite{Strom21}. % or constant rate factors (crfs) \cite{x264,x265}. %We choose QPs in such a way that the resulting RD points are located between the supporting points, such that they are located on the parts of the interpolated curves which are used for integration. 
Consequently, we encode bit streams with all available QPs between the minimum and the maximum considered QP. %An example for these points is shown in Fig.~\ref{fig:BD_example} with circle markers. 

To evaluate the accuracy of the interpolated RD-curves, we calculate the mean relative distance of the curve to all available points $p\in \mathcal{P}$ by 
\begin{equation}
\bar e = \frac{1}{K\cdot S\cdot \left|\mathcal{P}\right|}\sum_{k=1}^{K} \sum_{s=1}^{S} \sum_{p\in \mathcal{P}} \frac{\left| 10^{\hat r_{k,s}(D_{k,p,s})} - R_{k,p,s}\right|}{R_{k,p,s}},  \label{eq:MRE}
\end{equation}
where $k$ denotes the codec index with the number of codecs $K$ and $s$ denotes the sequence index with the total number of sequences $S$. 

\begin{figure}[t]
\centering
\providecommand\matlabtextA{\color[rgb]{0.150,0.150,0.150}}%
\psfrag{011}[tc][tc]{\matlabtextA Rate in Mbps}%
\psfrag{012}[bc][bc]{\matlabtextA PSNR in dB}%
\providecommand\matlabtextB{\color[rgb]{0.150,0.150,0.150}}%
\psfrag{000}[ct][ct]{\matlabtextB $0$}%
\psfrag{001}[ct][ct]{\matlabtextB $5000$}%
\psfrag{002}[ct][ct]{\matlabtextB $10000$}%
\psfrag{003}[ct][ct]{\matlabtextB $15000$}%
\psfrag{004}[rc][rc]{\matlabtextB $34$}%
\psfrag{005}[rc][rc]{\matlabtextB $35$}%
\psfrag{006}[rc][rc]{\matlabtextB $36$}%
\psfrag{007}[rc][rc]{\matlabtextB $37$}%
\psfrag{008}[rc][rc]{\matlabtextB $38$}%
\psfrag{009}[rc][rc]{\matlabtextB $39$}%
\psfrag{010}[rc][rc]{\matlabtextB $40$}%
\psfrag{Supporting Pointsssssssssss}[l][l]{Supporting points $R_{k,i,s}$}%
\psfrag{Additional points}[l][l]{Additional points $R_{k,p,s}$}%
\psfrag{Lin. Interp.}[l][l]{Interpolation $10^{\hat r_{k,s}(D_{k,p,s})}$}%
\psfrag{Interp. Error}[l][l]{Interpolation error}%
\includegraphics[width=.45\textwidth]{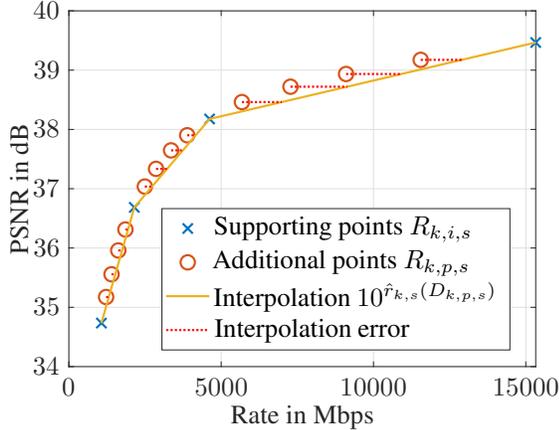} 
\vspace{-.4cm}
\caption{Supporting points (x'es), additional points (o's), a linearly interpolated curve (yellow), and the interpolation error (red dots). }
\label{fig:err_example}
\vspace{-.4cm}
\end{figure}

The calculus is visualized in Fig.~\ref{fig:err_example}. 
For illustration, we choose a fictional linear interpolation method.  %to visualize the distance between the true points $R_{k,p,s}$ and the interpolation $10^{\hat r_{k,s}(D_{k,p,s})}$. 
The supporting points and additional points are indicated by x'es and o's, respectively, and the interpolated curve is indicated by a yellow line. The interpolation errors for each point are calculated using the horizontal distance between the additional points $R_{k,p,s}$ and the interpolation $10^{\hat r_{k,s}(D_{k,p,s})}$, as indicated by the red dotted lines. 

Note that in this calculus, %we do not use the logarithmic representation of the rate because 
averaging is performed element-wise.  All elements of $\left|\mathcal{P}\right|$, i.e., all points in the region of interest, are distributed evenly between the four supporting points. As such, taking the logarithm of the rate \eqref{eq:logR} to avoid a bias towards higher bitrates is not required. %averaged over the number of points $\left|\mathcal{P}\right|$. Due to the definition of the QP, these points are already distributed in a logarithmic way, as can also be seen in Fig.~\ref{fig:BD_example}.
  
Intuitively, the relative error $\bar e$ describes the mean relative distance of the interpolated curves to the available RD points. As it is calculated in the same domain as the BD-rate, $\bar e$ can directly be interpreted as the maximum error imposed on the calculated BD-rate \eqref{eq:bdRateInt}, which is caused by interpolation. 

Next to the mean error, we also evaluate the maximum relative error 
\begin{equation}
E_\mathrm{max} = \max_{k, s, p} \frac{\left| 10^{\hat r_{k,s}(D_{k,p,s})} - R_{k,p,s}\right|}{R_{k,p,s}}, \label{eq:maxE}
\end{equation}
which corresponds to the largest relative horizontal distance of the interpolated to curve to the true RD points. % in the logarithmic domain. 
%This value provides additional information on outliers. 

\section{Evaluation Setup}
\label{sec:eval}

\begin{table}[t]
\renewcommand{\arraystretch}{1.3}
\caption{Evaluation sequences. }
\label{tab:sequences}
\vspace{-0.3cm}
\begin{center}
\begin{tabular}{l||c|c|c}
\hline
Name & Class & Resolution & Frame rate [fps]\\
\hline
Cactus & B & $1920\times 1080$ & 50 \\
PartyScene & C & $832\times 480$ & 50 \\
BQSquare & D & $416\times 240$ & 60 \\
BasketballPass & D & $416\times 240$ & 50 \\
FourPeople & E & $1280\times 720$ & 60 \\
SlideEditing & F& $1280\times 720$ & 30 \\
\hline
 \end{tabular}
\end{center}
\vspace{-.8cm}
\end{table}

\begin{table*}[t]
\renewcommand{\arraystretch}{1.3}
\caption{Mean relative errors and maximum relative errors for the five tested interpolation algorithms. }%VPORT - Bitrate is not shown because interpolation returns no results. 
%Best results are highlighted. } %The mean errors are averaged over all sequences and codecs, the maximum error is the maximum over all sequences, codecs, and QPs. NaN stands for not a number. }
\label{tab:errors}
%\vspace{-.35cm}
\begin{center}
\begin{tabular}{r||r|r||r|r||r|r||r|r||r|r}
\hline
% & \multicolumn{2}{c||}{CSI} & \multicolumn{2}{c||}{PCHIP} & \multicolumn{2}{c||}{Akima } &  \multicolumn{2}{c||}{CSI natural} & \multicolumn{2}{c}{CSI clamped}\\
 PM & \multicolumn{2}{c||}{PSNR - Bitrate} & \multicolumn{2}{c||}{SSIM - Bitrate} & \multicolumn{2}{c||}{VMAF - Bitrate} &  \multicolumn{2}{c||}{PSNR - Energy} & \multicolumn{2}{c}{VMAF - Energy}\\
pair & $\bar e $ & $E_\mathrm{max}$ & $\bar e$ & $E_\mathrm{max}$ & $\bar e$ & $E_\mathrm{max}$ & $\bar e$ & $E_\mathrm{max}$ & $\bar e$ & $E_\mathrm{max}$  \\
\hline

%    
%    CSI & $0.630\%$ & $5.151\%$  & $9.130\%$ & $110.446\%$  & $5.587\%$ & $29.093\%$  & $0.992\%$ & $\mathbf{7.060\%}$  & $2.588\%$ & $14.705\%$ \\ 
%     PHIP & $0.420\%$ & $4.103\%$  & $1.709\%$ & $9.329\%$  & $2.010\%$ & $12.971\%$  & $0.917\%$ & $7.093\%$  & $1.140\%$ & $\mathbf{7.046\%}$ \\ 
%     Akima & $\mathbf{0.370\%}$ & $4.855\%$  & $\mathbf{1.121\%}$ & $\mathbf{7.439\%}$  & $\mathbf{1.402\%}$ & $\mathbf{10.576\%}$  & $\mathbf{0.904\%}$ & $7.066\%$  & $\mathbf{1.064\%}$ & $7.053\%$ \\ 
%     CSI natural & $0.658\%$ & $\mathbf{4.246\%}$  & $3.870\%$ & $15.561\%$  & $3.450\%$ & $15.735\%$  & $1.019\%$ & $7.080\%$  & $1.765\%$ & $8.374\%$ \\ 
%     CSI clamped & $5.068\%$ & $25.984\%$  & $9.372\%$ & $44.081\%$  & $8.831\%$ & $37.198\%$  & $1.908\%$ & $12.882\%$  & $3.378\%$ & $15.558\%$ \\ 
%     
     
     CSI & \small $0.630\%$ & \small $5.151\%$  & \small $9.130\%$ & \small $110.446\%$  & \small $5.587\%$ & \small $29.093\%$  & \small $0.992\%$ & \small $\mathbf{7.060\%}$  & \small $2.588\%$ & \small $14.705\%$ \\ 
     PHIP & \small $0.420\%$ & \small $\mathbf{4.103\%}$  & \small $1.709\%$ & \small $9.329\%$  & \small $2.010\%$ & \small $12.971\%$  & \small $0.917\%$ & \small $7.093\%$  & \small $1.140\%$ & \small $\mathbf{7.046\%}$ \\ 
     Akima & \small $\mathbf{0.370\%}$ & \small $4.855\%$  & \small $\mathbf{1.121\%}$ & \small $\mathbf{7.439\%}$  & \small $\mathbf{1.402\%}$ & \small $\mathbf{10.576\%}$  & \small $\mathbf{0.904\%}$ & \small $7.066\%$  & \small $\mathbf{1.064\%}$ & \small $7.053\%$ \\ 
     \small CSI nat. & \small $0.658\%$ & \small $4.246\%$  & \small $3.870\%$ & \small $15.561\%$  & \small $3.450\%$ & \small $15.735\%$  & \small $1.019\%$ & \small $7.080\%$  & \small $1.765\%$ & \small $8.374\%$ \\ 
     \small CSI cla. & \small $5.068\%$ & \small $25.984\%$  & \small $9.372\%$ & \small $44.081\%$  & \small $8.831\%$ & \small $37.198\%$  & \small $1.908\%$ & \small $12.882\%$  & \small $3.378\%$ & \small $15.558\%$ \\ 
     
    \hline
 \end{tabular}
\end{center}
\vspace{-.8cm}
\end{table*}

%\subsection{Evaluation Setup}
%\label{secsec:dataSet}
For our evaluation, we use $S=6$ representative sequences from the JVET common test conditions \cite{JVET_N1010} as listed in Table~\ref{tab:sequences}. The sequences are provided with different resolutions and frame rates. We encode ten seconds using the QPs $\in\{\mathbf{22}, 23, 24, .., \mathbf{27}, .., \mathbf{32}, .., \mathbf{37}\}$ to generate the PM points in $\mathcal{P}$, where the bold values are used as supporting points for the calculation of the interpolation, as proposed by the HEVC and the VVC common test conditions \cite{Bossen13,JVET_N1010}. 

 As encoders, we choose the HM encoder and the VTM encoder for HEVC and VVC, respectively, ($K=2$ codecs). As a configuration, we choose random access with an internal bit depth of $10\,$bit for both encoders \cite{Bossen13,JVET_N1010}. 

We evaluate the following interpolation methods. First, we use cubic spline interpolation  (CSI) with a not-a-knot boundary constraint, which was initially used for BD calculations \cite{Tan16} and requires the first and the second derivative to be continuous. % and corresponds to \eqref{eq:BDpoly}. 
Second, we evaluate CSI with a natural boundary constraint, which means that the second derivative at the two ends of the interpolation curve is set to zero (CSI - natural). Third, we replace the boundary constraint with a clamped constraint, meaning that the derivative at the borders is zero (CSI - clamped). Fourth, we evaluate piecewise-cubic hermite interpolating polynomial (PCHIP), which was found to be more stable for special cases of RD curves \cite{Zhao08} and which is also used in practical BD calculations for, e.g., standardization \cite{Strom21}. Finally, we test Akima interpolation \cite{Akima70}, which  returns a piecewise polynomial, too. The latter two methods only require the first derivative of the polynomial to be continuous. % that is continuously differentiable only once.  % and third, we discuss the least-squares polynomial fit (LSPF).  
%In latest standardization activities, it

%\subsection{Performance Metrics}
%\label{secsec:perfMetrics}
As performance metrics, next to the traditional bitrate and PSNR, we choose to evaluate the quality metrics SSIM \cite{Wang04} and VMAF \cite{VMAF}. %, where we use two models: standard VMAF and VMAF trained for portable devices (VPORT). 
These two metrics replace the PSNR describing the distortion $D$. Concerning an alternative for the bitrate $R$, we consider the decoding energy of software decoders \cite{Herglotz18}, where we perform VTM and HM decoding on an Intel processor and measure the decoding energy with running average power limit (RAPL) \cite{David10}. Furthermore, one PM pair is chosen where both traditional RD-metrics are replaced with other metrics (VMAF - decoding energy). In total, we evaluate five PM pairs.  %All considered PM pairs and example curves are illustrated in Fig.~\ref{fig:PMcurves}. 

\section{Accuracy of Interpolation Curves}
\label{secsec:evalInterp}

\begin{figure}[t]
\centering
\providecommand\matlabtextA{\color[rgb]{0.150,0.150,0.150}}%
\psfrag{009}[c][c]{\matlabtextA Rate in Mbps}%
\psfrag{010}[bc][bc]{\matlabtextA SSIM}%
\providecommand\matlabtextB{\color[rgb]{0.150,0.150,0.150}}%
\psfrag{000}[ct][ct]{\matlabtextB $1$}%
\psfrag{001}[ct][ct]{\matlabtextB $10$}%
\psfrag{002}[rc][rc]{\matlabtextB $0.97$}%
\psfrag{003}[rc][rc]{\matlabtextB $ $}%
\psfrag{004}[rc][rc]{\matlabtextB $0.98$}%
\psfrag{005}[rc][rc]{\matlabtextB $ $}%
\psfrag{006}[rc][rc]{\matlabtextB $0.99$}%
\psfrag{007}[rc][rc]{\matlabtextB $ $}%
\psfrag{008}[rc][rc]{\matlabtextB $1$}%
\psfrag{Supporting points}[l][l]{\small Supporting points}
\psfrag{Additional points}[l][l]{\small Additional points}
\psfrag{CSI}[l][l]{\small CSI}
\psfrag{PCHIP}[l][l]{\small PCHIP}
\psfrag{Akima}[l][l]{\small Akima (\newBDnameShort)}
\includegraphics[width=.45\textwidth]{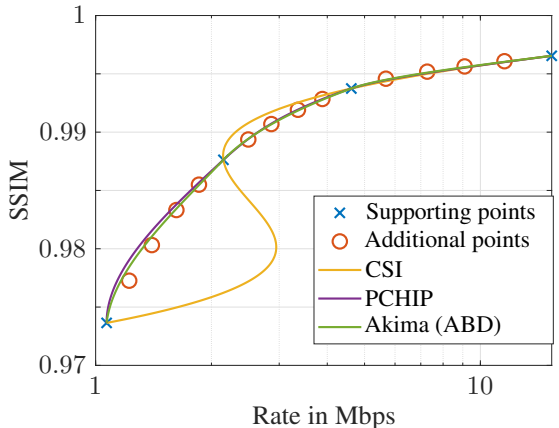} 
\vspace{-.5cm}
\caption{Supporting points (x'es), additional points (o's), and the curves interpolated by CSI (yellow), PCHIP (purple), and Akima (green) for the Cactus sequence (SSIM - bitrate).  }
\label{fig:Cactus_Bad}
\vspace{-.2cm}
\end{figure}

For illustration of the performance of the interpolation methods, in Fig.~\ref{fig:Cactus_Bad}, we plot supporting points, additional points, and example interpolated curves for the VVC-coded Cactus sequence, which returned the overall highest maximum interpolation error. The PM pair is SSIM - Bitrate. The curve is plotted in a logarithmic representation to highlight differences between the interpolated curves. 

Apparently, the interpolation returned by CSI leads to an overshoot of the interpolated curve, which causes the interpolated bitrate to be more than twice as high as measured (the marker close to an SSIM value of $0.98$). In contrast, the PCHIP algorithm returns a stable curve with a maximum error below $10\%$. Still, Akima interpolation returns the closest approximation (visible at low bitrates) as the green curve is located closer to the markers than the other curves.

The resulting mean relative errors $\bar e$ and maximum errors $E_\mathrm{max}$ for all PM pairs are listed in Table~\ref{tab:errors}. 
The results confirm observations in \cite{Strom21}, stating that PCHIP returns more stable interpolations than CSI because in some cases (SSIM - Bitrate, VMAF - Bitrate, and VMAF - Energy), the mean relative error is more than twice as high as for PCHIP. However, we can also see that Akima interpolation outperforms all the other interpolation methods for all PM pairs. Considering the errors of CSI natural and CSI clamped, we can see that these interpolation methods are not well suited for BD calculations. 

In general, we observe mean interpolation errors below $1.5\%$ for the Akima method. When comparing different codecs, which typically show BD differences significantly larger than $5\%$, this accuracy is sufficient. However, in the development of new coding tools, BD differences often yield values below $1\%$. For such cases, future research could investigate whether the interpolation errors shown in Table~\ref{tab:errors} can lead to incorrect decisions on the adoption of coding tools. 

Considering the maximum relative error, we can see that Akima interpolation does not always lead to the most accurate interpolation. However, the difference to the best result is always lower than $0.7\%$, such that we conclude that Akima interpolation is the most accurate way to interpolate PM curves. For practical use, we provide Python scripts for cubic, piecewise cubic, and Akima interpolation at \cite{ABD}.

\section{Conclusions}
\label{sec:concl}
In this paper, we evaluated five interpolation algorithms of the Bj{\o}ntegaard-Delta calculus in detail using two different video codecs and five pairs of performance metrics.  The results suggest that a novel method based on Akima interpolation (\newBDnameShort) returns most accurate interpolations for different performance metric curves. Furthermore, we could show that when comparing the compression performance of different codecs, useful results can be obtained when the BD-rate difference is larger than $1.5\%$. %The results also indicate that the BD calculus cannot be used when the quality metric saturates. 

In future work, %a feasible solution for saturated supporting points could be developed. Furthermore, 
the same evaluations could be performed for further performance metrics such as subjective quality scores or mean average precision scores in object detection algorithms. %Furthermore, the impact of the interpolation error on the final BD-rate value could be investigated in more detail.  %Finally, it would be interesting to investigate the impact of the errors when two PM curves are very close. 

\let\oldthebibliography=\thebibliography
\let\endoldthebibliography=\endthebibliography
\renewenvironment{thebibliography}[1]{%
   \begin{oldthebibliography}{#1}%
     \setlength{\itemsep}{-.4ex}%
}%
{%
   \end{oldthebibliography}%
}

\end{document}